\documentclass[bibnotes,floats,aps,a4paper,twocolumn,showpacs,superscriptaddress]{revtex4}
\usepackage{epsfig}
\begin{document}

\title{Doppler Effect of Nonlinear Waves and Superspirals 
in Oscillatory Media}
\author{Lutz Brusch}
\affiliation{MPI for Physics of Complex Systems, 
 N\"othnitzer Str. 38, D-01187 Dresden, Germany}
\email{baer@mpipks-dresden.mpg.de}
\author{Alessandro Torcini}
\affiliation{Istituto Nazionale di Ottica Applicata, 
L.go E. Fermi 6, I-50125 Firenze, Italy}
\email{torcini@ino.it}
\author{Markus B\"ar}
\affiliation{MPI for Physics of Complex Systems, 
 N\"othnitzer Str. 38, D-01187 Dresden, Germany}
\date{\today} 

\begin{abstract}
Nonlinear waves emitted from a moving source are studied. 
A meandering spiral in a reaction-diffusion medium provides an 
example, where waves originate from a source exhibiting a back-and-forth
movement in radial direction.
The periodic motion of the source induces a Doppler effect that causes a 
modulation in wavelength and amplitude of the waves (``superspiral'').
Using the complex Ginzburg-Landau equation,
we show that waves subject to 
a convective Eckhaus instability can exhibit 
monotonous growth or decay as well as saturation
of these modulations away from the source
depending on the perturbation frequency. 
Our findings allow a consistent interpretation of recent experimental
observations concerning superspirals and their 
decay to spatio-temporal chaos.
\end{abstract}


\maketitle

{\em Introduction. -} 
Periodic nonlinear waves are a trademark of nonequilibrium  systems 
\cite{general}. 
In one dimension (1D), they can appear in systems with periodic
boundary conditions (BCs) as well as in open geometries, where 
BCs select a unique pattern \cite{hagan82}. 
In two dimensions (2D), 
rotating spiral waves are frequently observed. 
Therein, periodic waves emerge from the region of the spiral tip 
(core) and propagate in radial direction. 
The aim of this paper is to investigate
the effects of perturbing 
sources of nonlinear waves and their implications for 
the dynamics of spiral waves. 
We employ the complex Ginzburg-Landau equation (CGLE), 
which provides an
universal description of spatially extended oscillatory systems 
near a supercritical Hopf bifurcation \cite{kuramoto,lkrmp02}. 

In this framework,
we perturb a source of periodic waves in 1D by moving its position
back-and-forth in space. 
This motion of the source leads  to a modulation in 
amplitude, wavelength and frequency of the emitted waves. 
We find that the modulation of the nonlinear waves is uniquely 
determined by the temporal period of the source motion 
in contrast to linear waves emitted by a moving source, where the 
source velocity is the relevant quantity. 
The richest scenario is found if the emitted waves 
are convectively Eckhaus unstable.
If we consider a periodically moving source in the
latter case,
the modulations in amplitude and period of the waves may   
(a) get exponentially damped, (b) saturate  
or (c) grow monotonously far away from the source 
depending on the frequency of the applied forcing. 

\begin{figure}
\begin{center}
 \epsfxsize=1.0\hsize
 \epsffile{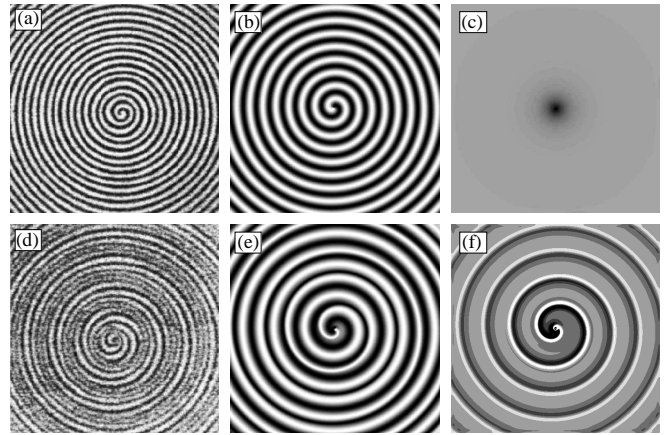}
\end{center}
 \caption
 {Simple spiral waves (a-c) and superspirals (d-f) are
 observed in experiments (a,d) of the BZ system \cite{qi01} and 
 in numerical simulations of the CGLE (\ref{cgle1d}).
 (b,e) show Re$[A]$ and (c,f) $|A|$.
 (b,c) $\mu=1$ and (e,f) inhomogeneous $\mu=1+0.7 \exp(-r^2/10)$.
 Other parameters are $c_1=3.5, c_3=0.34$, system size is $512\times 512$.
}
 \label{suspir}
\end{figure}
Such a periodically moving source reproduces
the radial dynamics of rotating spiral waves subject 
to external forcing \cite{extforc} or 
to the frequently observed oscillatory instability 
termed ``meandering'' \cite{meander}. 
In 2D, the radial modulations caused by meandering lead to 
a second spiral superimposed on the simple rotating spiral;
the resulting structure is called superspiral (see Fig. 1). 
Our study provides insight in the nonlinear behavior of such
superspirals extending earlier work based on a linear analysis~\cite{ssprl01}.
Unbounded
growth of the modulation will lead to occurrence of space-time defects 
in 1D and superspiral breakup in 2D, provided the system is sufficiently large.
Experimental evidence of such behavior was found in the
Belousov-Zhabotinsky (BZ) reaction \cite{qi01}.
The first experimental observation of a superspiral 
was reported for a spiral subject to strong external forcing
near its core \cite{krinsky91}. 

{\em Complex Ginzburg-Landau equation. -} 
Consider a spatially extended oscillatory medium described by the 
complex Ginzburg-Landau equation \cite{general,lkrmp02}
\begin{equation} 
  \partial_{t} A = \mu A + (1+ ic_1) \Delta A - (1-i c_3) |A|^2 A
\label{cgle1d}
\end{equation}
with $\mu=1$. 
The complex field $A(r,t)$
gives amplitude and phase of local oscillations depending on real
coefficients $c_1, c_3$ determined by the underlying 
specific model at the onset of oscillations.  
Eq. (\ref{cgle1d}) exhibits plane wave solutions  
$ A(r,t) = \sqrt{1-q^2} \mbox{e}^{i (q r -
\omega t)} $ with $\omega = -c_3+q^2(c_1+c_3)$
in an infinite or periodic 1D medium with $\Delta=\partial_{r}^{2}$.
For fixed control parameters, these waves become unstable
if the wavenumber $q$ is bigger than the Eckhaus wavenumber 
$q_E^2 = (1 - c_1 c_3)/(2 (1 + c_3^2) + 1 - c_1
c_3)$. 
The plane waves represent a one-parameter family that is 
parametrized by the wavenumber $q$. 
A source of periodic waves at $r=0$ is 
easily realized by applying the BCs 
\begin{equation}
A(r=0) = 0 \quad \mbox{and} \quad  \partial_r A (r=L) =
0. 
\label{dirichlet_bc}
\end{equation}
The specification of these BCs  
leads to the selection of a unique wavenumber $q_S$. 
In this case, 
the exact solution of the CGLE is of the form
\begin{equation}
A(r,t) = F(r) \, \mbox{e}^{\displaystyle i(f(r)-\omega t)},
\label{spiralform}
\end{equation}
with the asymptotic behavior 
$df(r)/dr \to q_S, F(r) \to \sqrt{1-q_S^2}$ for $r\to\infty$ (far-field) and 
$df(r)/dr \sim r, F(r) \sim r$ for $r\to 0$ (near-field).
An analytic expression for $q_S$ has been derived
\cite{hagan82,lkbook90}
\begin{equation} 
0 = (c_1 +c_3) q_S^2 + 3 \alpha(c_1,c_3) q_S -c_3 -2 c_1
\alpha(c_1,c_3)^2,
\label{qhagan}
\end{equation}
where $\alpha$ is a function of the control parameters
$c_1, c_3$ \cite{footnote}. 

In 2D, Eq. (\ref{cgle1d}) with
$\Delta=\partial_{r}^{2}+1/r \partial_r + 1/r^2 \partial_{\theta}^2$
possesses rotating spiral solutions 
with $A=0$ in the spiral core and 
in the far-field a selected wavenumber $q_S$,
which is similar to (4) as verified numerically \cite{hagan82,lkbook90}.
Eq. (4) allows to 
discriminate if the source or the spiral core emits 
stable (Eckhaus unstable) wavetrains with $q_S < q_E$
($q_S > q_E$).
Further analysis showed that the Eckhaus instability is of convective 
nature and becomes absolute 
for sufficiently large 
values of $c_1$ and $c_3$ \cite{lkpra92}. 
At given parameters, we may define a wavenumber $q_A > q_E$ 
that characterizes
absolutely unstable waves with $q > q_A$. 
In finite systems, spirals were found to be stable 
as long as $q_S < q_A$ \cite{lkpra92,stprl98}. 
\begin{figure}
\begin{center}
 \epsfxsize=1.0\hsize 
 \epsffile{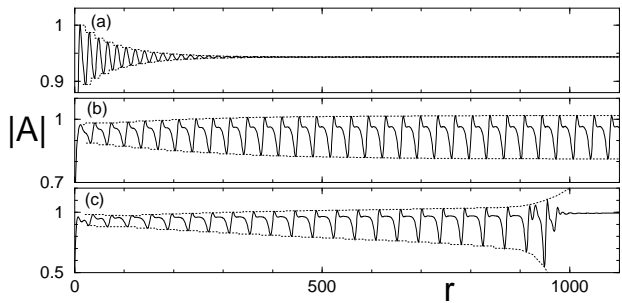}
\end{center}
 \caption{
 Doppler effect of nonlinear waves due to periodic oscillations of 
 the source at the left boundary.
 Eq. (\ref{cgle1d}) was integrated for increasing forcing periods
 (a) $\tau=8$, (b) $\tau=13$, (c) $\tau=15$
 with Eq. (\ref{osc_bc}) and $c_1=3.5, c_3=0.4, R_S=1$.
 Dotted lines are guides to the eye that converge to $|A|_{min}$ and 
$|A|_{max}$  in (b).
 }
 \label{profiles}
\end{figure}

{\em Simulation results. -} 
Here, we analyze the effects of an instability
or of an external perturbation at the source 
or at the spiral core (meandering).
We choose to perturb the source in Eq. (\ref{dirichlet_bc})
by varying its position $r_S$.
Consequently,
\begin{eqnarray}
A(r \le r_S) = 0  \quad \mbox{with} \quad r_S=R_S \cos (2\pi t/\tau)  
\nonumber \\ 
\quad \mbox{and} \quad \partial_r A (r=L) =0, 
\label{osc_bc}
\end{eqnarray}
is used as BC of Eq. (\ref{cgle1d}) in 1D.
The core motion of a meandering spiral can be viewed as a 
source moving along a circle in 2D. 
A projection onto a radial direction for fixed angle yields a periodic 
back-and-forth motion.
A sinusoidal motion like in Eq. (5) is expected near onset, where the 
normal form of meandering \cite{dbprl94} provides a valid description.

Hence, the BC in Eq. (5) also captures the radial dynamics 
of a meandering spiral.
In the numerical simulations the moving source initially modulates the 
local
wavenumber which subsequently leads to a modulation of the wave amplitude. 
For large $r$ (far-field),  we observe modulations 
with a period $T$ that is equal to the forcing period $\tau$
and independent from  $R_S$.
First, we forced sources (respectively ``spirals'') in the 
parameter region where $q_S <
q_E$, in this case the perturbation is always damped out while the
waves move away from the source. 
A larger variety of responses occurs in the convectively Eckhaus unstable
regime, where the unperturbed source selects a wavenumber $q_S$ 
in the interval $[q_E,q_A]$.
Fig. 2 shows three qualitatively different 
resulting profiles of the amplitude $|A|$ 
as a function of the radial coordinate. 
Figs. 2(a), (b) and (c) correspond to increasing values of
the forcing period $\tau$.
For small $\tau$-values, 
the result is similar to the case with $q_S
< q_E$, the amplitude modulation is exponentially damped and is barely
visible at sufficiently large $r$ (Fig. 2(a)). 
For intermediate $\tau$, the amplitude modulation first grows and then
reaches saturation (Fig. 2(b)). 
The profile of $|A|$ is periodic and travels with a non-zero velocity; 
in the far-field, the radial dynamics resemble a so-called modulated 
amplitude wave (see below).
Finally for large $\tau$, the amplitude modulation grows monotonically
with $r$ and space-time defects are formed at the breakup radius 
$r=R_{BU} \approx 950$  (Fig. 2(c)).
For simulations with $q_S > q_A$, we always observe monotonous growth of
the modulation leading to space-time defects. 

A drawback of the homogeneous CGLE in 2D is that
it does not support meandering spirals. 
A non-saturating meandering instability has been
observed for large values of $c_1$ \cite{lkprl94}. 
Alternatively, addition of a heterogeneity near the
spiral core leads to meandering behavior similar to the 
one typically seen in reaction-diffusion systems \cite{lkbook95}. 
%
Figs. 1 (e),(f) show such a superspiral.
The corresponding quantities for a regular spiral are shown in
Fig. 1 (b),(c). 
As in Fig. 2(b) and in the recent experiment 
by Zhou and Ouyang (see Fig. 1(d) and \cite{qi01}) 
the amplitude modulation shown in
Fig. 1(f) saturates in the far-field.
In the following, the study will be limited to the radial dynamics.
One of the new results of our analysis is that the modulation 
of the amplitude may eventually saturate (see Fig. 2(b)). 
It also travels with constant shape and with a speed
different from that of the underlying phase waves. 
To summarize, 
periodically moving sources emit modulated amplitude waves 
analogous to the way stationary sources send out plane waves.  

\begin{figure}
\begin{center}
 \epsfxsize=1.0\hsize 
 \epsffile{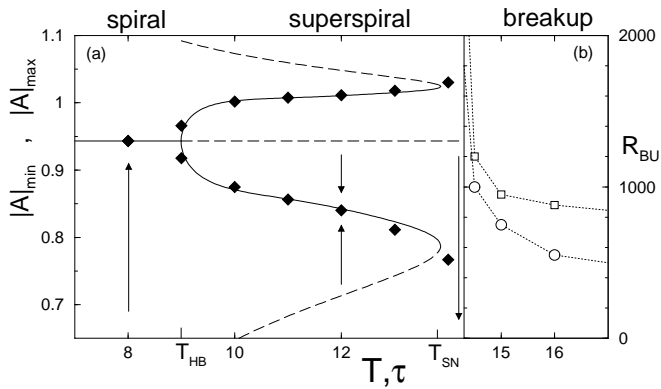}
\end{center}
 \caption
 {(a) Bifurcation diagram for MAWs with period $T$ 
 and existence domains of spirals, superspirals and spiral breakup for various
forcing periods $\tau$. Curves were computed as in \cite{lbmaw,lbmaw2} and
symbols denote minima and maxima of $|A|$ as measured in 
simulations, parameters are the same as in Fig. 2. 
Solid (dashed) curves correspond to stable (unstable) solutions.
Arrows (same for $|A|_{max}$) indicate the evolution of initial 
perturbations as they move away from
the source. Their asymptotic values depend on $T=\tau$ but not on $R_S$.
(b) Breakup radius $R_{BU}$ at which defects firstly occur in
simulations. Smaller remaining spirals result from larger $R_S$ ($R_S=5$
circles, $R_S=1$ squares).
 }
\label{Tbif}
\end{figure}

{\em Modulated amplitude waves (MAWs) and  superspirals. -} 
MAWs are solutions of the CGLE and have 
the following general form
\begin{equation} 
A(r,t) = a(z) e^{i\phi(z)}  e^{i(q r-\omega t)},
\label{maw}
\end{equation}
with unknown periodic functions $a$ and $\phi$ of the comoving coordinate 
$z = r - v t$ \cite{lbmaw}. 
The analysis of MAWs has revealed that they bifurcate from plane 
waves with $q > q_E$. 
MAWs form a two-parameter family of waves described by the wavelength $P$ 
of the modulation of the amplitude $a = |A|$ and by the spatially averaged phase 
gradient $\nu=q+\langle \partial_r \phi \rangle$.
The velocity $v$ and the period $T=P/v$ can be derived 
from these quantities. 
The average phase gradient is conserved as long as no space-time
defect is formed. 
For periodically perturbed sources, the phase gradient value is
fixed to $\nu \equiv q_S$. 
Therefore, only one free parameter is left.
For the present purpose it is most convenient 
to use the period $T$.
The resulting bifurcation diagram is shown in Fig. 3 for the case $q_S
> q_E$ (Eckhaus unstable range). 
In the Eckhaus stable regime, no stable MAWs exist. 

Fig. 3 reveals that stable MAWs exist 
for  periods $T$ in the interval $[T_{HB},T_{SN}]$.
They are ``born'' in a Hopf bifurcation (HB) and ``die'' in a saddle-node
bifurcation (SN).
For $T < T_{HB}$, the plane wave is stable and for $T > T_{SN}$ no
MAWs exists and the dynamics leads to the formation of 
defects~\cite{lbmaw,vanhecke}.
This clarifies our findings in Fig. 2: the profiles in Fig. 2(a), 2(b)
and 2(c) correspond to forcing within the regions $\tau < T_{HB}$, 
$T_{HB} < \tau <T_{SN}$ and $T_{SN} < \tau$, respectively. 
Fig. 3 (b) shows the breakup radius $R_{BU}$ found in simulations for
periods $\tau > T_{SN}$. 
It is crucial to realize that the properties of the saturated modulations, 
observed in numerical simulations with a periodically moving source, 
indeed correspond to the MAWs with $\nu = q_S$ and $T = \tau$ 
computed via a bifurcation analysis (as shown in Fig. 3).
Thus, a unique MAW characterized by two parameters 
is selected in 1D by BC in Eq. (5) or 
in 2D by the two intrinsic frequencies of a meandering spiral. 
We conjecture, that the bifurcation diagram of superspirals in 2D 
can be predicted from the corresponding bifurcation
diagram of the far-field MAWs provided the meandering period 
(that here corresponds to $\tau$) is known. 
Hence, the superspirals with saturated modulation 
should cease to exist in the saddle-node 
bifurcation of the associated MAW. 
If $\tau > T_{SN}$, the modulations grow monotonously and 
lead to appearance of space-time
defects in 1D, respectively topological defects in 2D 
(spiral breakup). 
In explicit 2D simulations, such behavior has already been
observed in a homogeneous reaction-diffusion model for calcium waves, 
where simultaneous appearance of the Eckhaus instability and 
meandering leads to a breakup far away from the spiral core \cite{calcium}.
Future studies may address the possibility of MAWs in realistic
reaction-diffusion systems which exhibit Eckhaus instabilities 
\cite{mbprl99,calcium}.
\begin{figure}
\begin{center}
 \epsfxsize=1.0\hsize 
 \epsffile{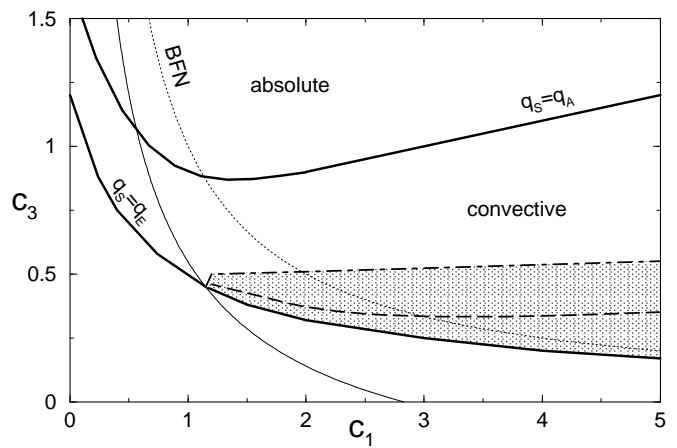}
\end{center}
 \caption
{Phase diagram indicating the region (shaded) where superspirals 
may occur in the CGLE for ``rocking'' sources.
Simple spirals with selected $q_S (c_1,c_3)$ are convectively unstable
between the thick solid curves.
Superspiral breakup is only possible between the dashed and dash-dotted
curves.
The Eckhaus instability for arbitrary $q$ is supercritical above the thin
solid curve and the thin dotted curve indicates the Benjamin-Feir-Newell line
(BFN).
 }
\label{c1c3}
\end{figure}
A previous linear analysis of superspirals
predicted that the meandering instability of a spiral 
with an Eckhaus unstable wavetrain in the far-field may produce 
superspirals with exponentially growing modulation, while in the
standard case of a meandering spiral emitting a stable wave train 
exponential damping of the modulation should be observed \cite{ssprl01}.
Our nonlinear analysis introduces superspirals with a saturated 
modulation in the far-field
as a third possibility, thus supporting the experimental observation
of such structures in the BZ reaction (Fig. 1 and \cite{qi01}).
We suggest an interpretation of 
the reported scenario damped superspiral - saturated 
superspiral - far-field breakup:
the three phenomenologies simply correspond to the three mentioned regions
of the bifurcation diagram for the MAWs. 
Notice that bifurcation diagrams similar to Fig. 3 (a) are found when 
$c_1$ or $c_3$ are varied at fixed $T$ \cite{lbmaw,lbmaw2,vanhecke}.

{\em Phase diagram. -} 
Finally we determine the regions of the $c_1-c_3$ phase diagram
\cite{lkpra92} 
where the discussed phenomena may be found. 
The results are summarized in Fig. 4 
where $\nu$ is fixed equal to $q_S(c_1, c_3)$ as given by Eq. (4).  
Saturated superspirals can appear only above the Eckhaus line 
for $q_S > q_E$.
For small values of $c_1$, the MAWs bifurcate subcritically and are
always unstable \cite{janiaud}. 
Fig. 4 also shows the line of absolute instability ($q_S = q_A$) above
which simple spirals break up giving rise to spatio-temporal chaos. 
Stable MAWs with $\nu=q_S$ do exist in the shaded region 
(for an extensive discussion on MAW stability see \cite{lbmaw2}). 
Below the thick dashed line in Fig. 4 no saddle-node bifurcation 
does occur and breakup is therefore prevented.
Between the thick dashed and the thick dash-dotted lines in Fig. 4, 
bifurcation diagrams similar to Fig. 3(a) are found. 
In this region, all the three behaviors reported in Fig. 2 are
possible, depending on the period of the forcing or meandering. 
Altogether, the results here presented link far-field breakup of
meandering spirals to a saddle-node bifurcation. 
Thus, this route to spatio-temporal chaos can be distinguished from  
the previously reported scenario of far-field breakup, 
where spiral breakup has been explained by an absolute Eckhaus 
instability of the asymptotic far-field 
wavetrain~\cite{lkpra92,stprl98,mbprl99,qinat96}.

{\em Conclusions. -} 
We have studied the Doppler effect 
associated with the back-and-forth motion 
of a source emitting periodic nonlinear waves. 
Usually, the resulting modulation dies out by exponential damping as the waves 
move away from the source. 
If the emitted wavetrain is convectively Eckhaus unstable, 
the wave modulation can also saturate or grow exponentially far away from the 
source depending on the period of the back-and-forth motion. 
These scenarios are fully determined by 
the bifurcation diagram of corresponding MAWs. 
Our results offer a consistent explanation of recent experimental
results obtained in a chemical reaction \cite{qi01}. 
Moreover, they may find application in future studies of other
experimental systems exhibiting convective instabilities 
like hydrothermal waves~\cite{hydrotherm}. 
Periodic forcing of sources near the transition
to chaotic or turbulent dynamics may be used to probe the 
existence and properties of modulated structures. 

{\em Acknowledgements} - We acknowledge stimulating discussion with 
B. Sandstede and Q. Ouyang, in addition we would like to thank Q. Ouyang
for the permission to report his experimental images in 
Fig. 1(a),(d). 



\begin{thebibliography}{99}
\bibitem{footnote} 
The function $\alpha$ is given by 
$\alpha(c_1,c_3) =
\sqrt{f(c_1, c_3)/g(c_1,c_3)}$ with 
$f(c_1,c_3) = 3 c_1 (8 (c_1-c_3)^2+9 (1+c_1 c_3)^2-4 c_1 c_3)^{1/2}
+c_1 (5-9 c_1 c_3) -4 c_3$ and  
$g(c_1,c_3) = 4 (-2 c_3+9 c_1^3+7 c_1)$.
\end{thebibliography}
\end{document}